\documentclass{cernyrep}
\begin{document}
\title{Plans for super-beams in Japan\footnote{\setlength{\baselineskip}{4mm}Contribution to the Workshop ``European Strategy for Future Neutrino Physics'', CERN, Oct. 2009, to appear in the Proceedings. Manuscript with high resolution figures is available at \texttt{http://jnusrv01.kek.jp/}$\sim$\texttt{hasegawa/pub/cernreport.pdf .}}}
\author{Takuya Hasegawa}
\institute{High Energy Accelerator Research Organization (KEK), Tsukuba, Ibaraki 305-0801, Japan}
\maketitle

\begin{abstract}
In Japan, as the first experiment utilizes J-PARC (Japan Proton Accelerator Research Complex) neutrino facility, 
T2K (Tokai to Kamioka Long Baseline Neutrino Experiment) starts operation. 
T2K is supposed to give critical information, which guides the future direction of the neutrino physics. 
Possible new generation discovery experiment based on T2K outcome is discussed. 
Especially, description of J-PARC neutrino beam upgrade plan and discussion on far detector options 
to maximize potential of the research are focused. European participation and CERN commitment on Japanese accelerator
based neutrino experiment is also reported.  
\end{abstract}

\section{J-PARC and Main Ring Synchrotron}

\begin{figure}[htb]
\begin{center}
\includegraphics[width=9cm,angle=-90]{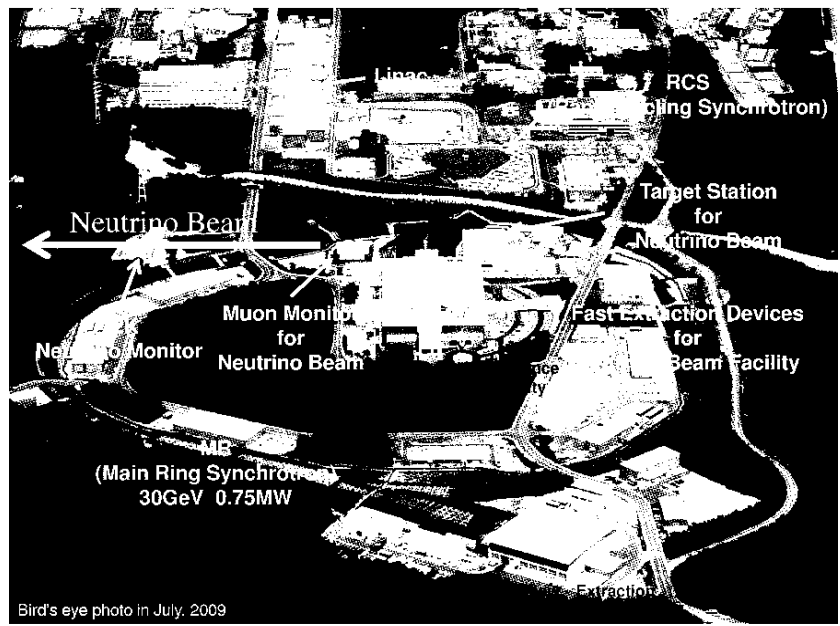}
\caption{J-PARC accelerator and experimental facility}
\label{fig:J-PARC}
\end{center}
\end{figure}

J-PARC (Figure~\ref{fig:J-PARC}) is a KEK-JAEA joint facility for MW-class high intensity proton accelerator.
It provides unprecedented high flux of various secondary particles, such as neutrons, muons, pions, kaons, and neutrinos,
which are utilized for elementary particle physics and material and life science.
%
%
%

\begin{figure}[htb]
\begin{center}
\includegraphics[width=9cm,angle=-90]{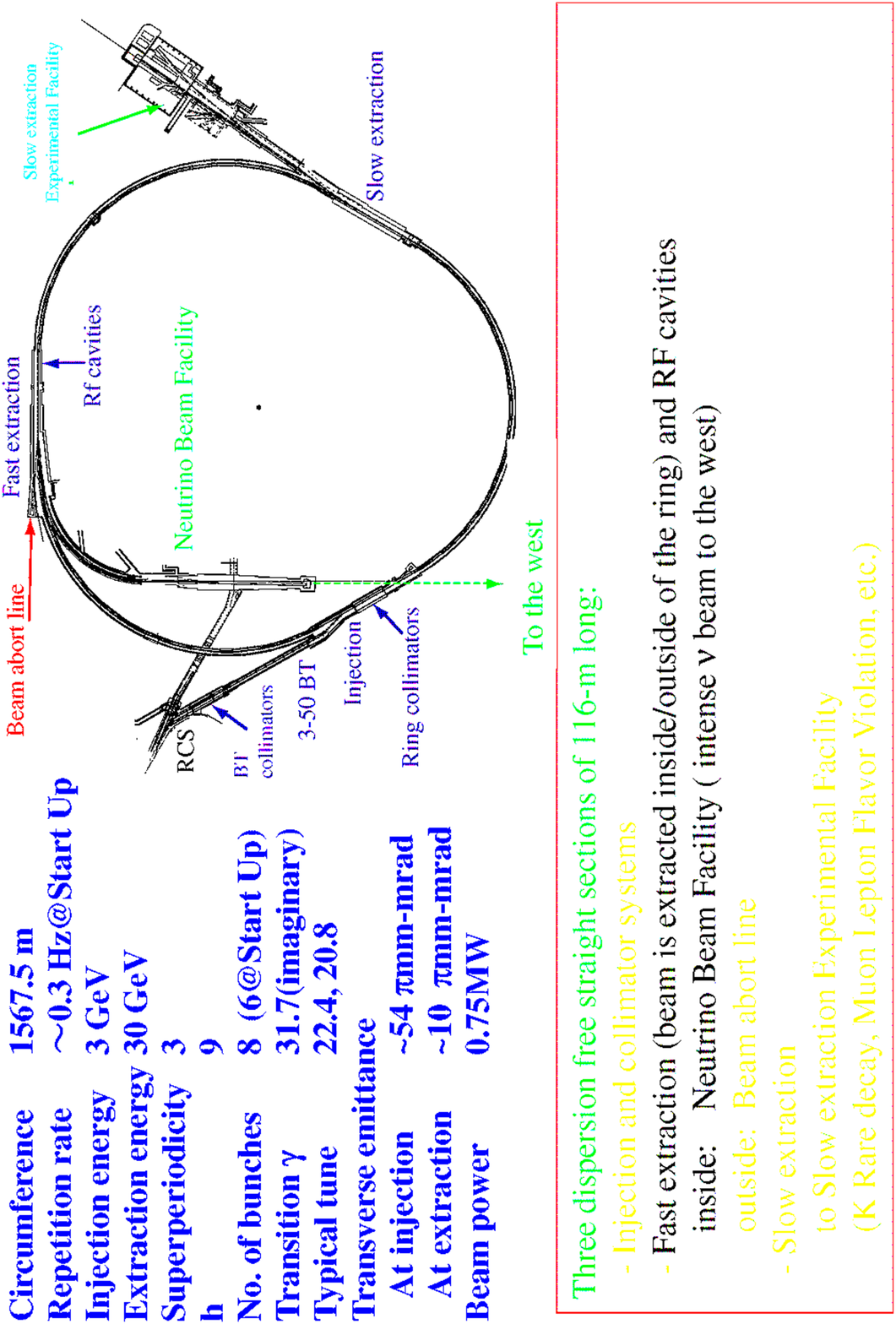}
\caption{Overview of MR}
\label{fig:MR}
\end{center}
\end{figure}

In the accelerator complex, H$^{-}$ ions are accelerated to 181~MeV with LINAC, fed
into Rapid Cycling Synchrotron (RCS) with stripping out electrons and are accelerated to 3~GeV. At final stage, proton beam goes into Main Ring Synchrotron (MR) 
and accelerated to 30~GeV. For the neutrino experiment, accelerated protons are
kicked inward to neutrino beam facility by single turn with fast extraction devices. 
Main characteristics of MR is described in Figure~\ref{fig:MR}.
%
%
%
%
%
%
%
%
\section{J-PARC neutrino beam facility}
\begin{figure}[htb]
\begin{center}
\includegraphics[width=9cm,angle=-90]{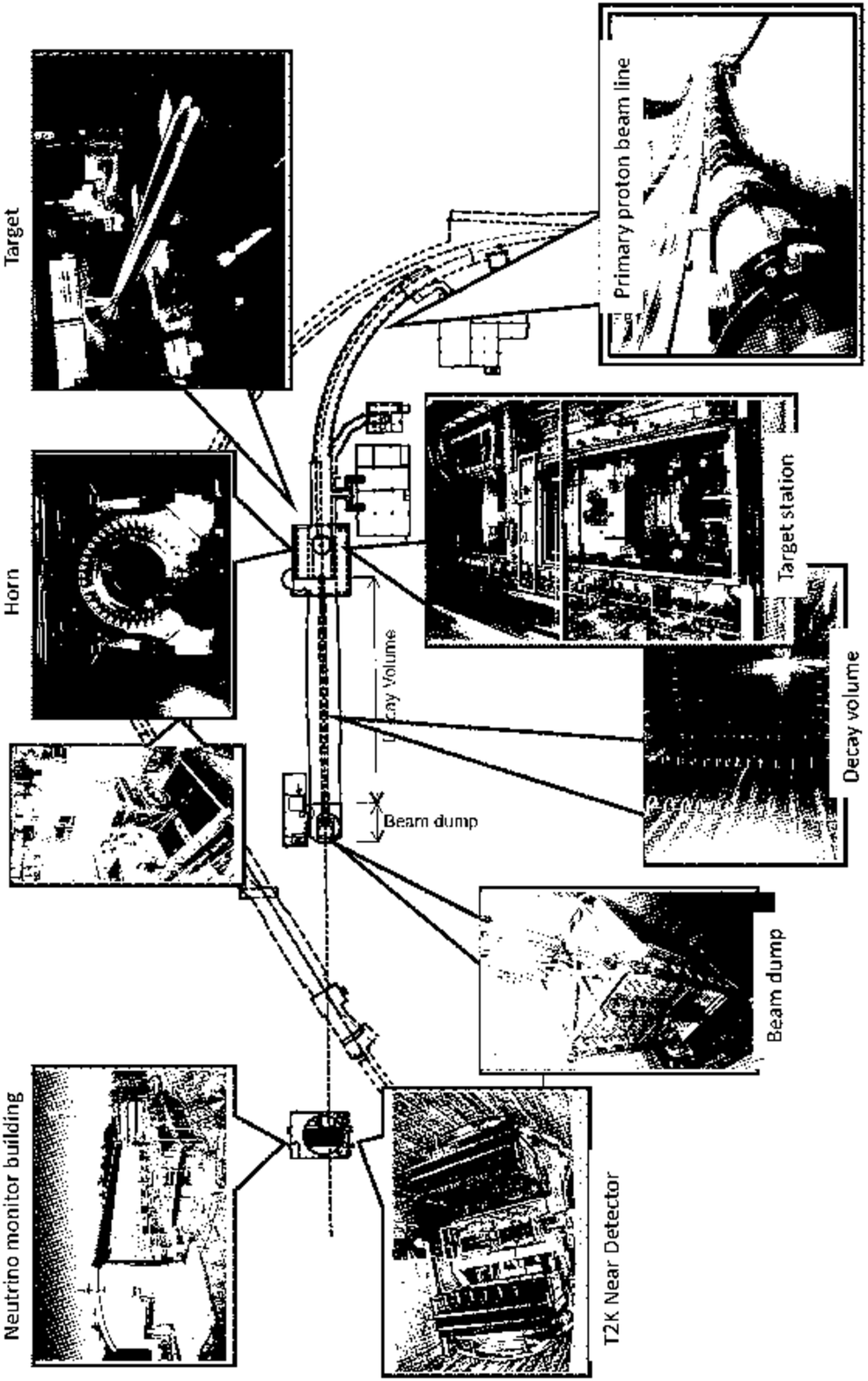}
\caption{J-PARC neutrino beam facility}
\label{fig:neutrino}
\end{center}
\end{figure}

%
%
The proton beam from MR run through J-PARC neutrino beam facility and producing
intense muon neutrinos toward the west direction. 
J-PARC neutrino beam facility is composed of following 
parts with their functionalities (Figure~\ref{fig:neutrino}).
\begin{itemize}
\item Preparation section: Match the beam optics to the arc section.
\item Arc section: Bend the beam $\sim$90$^\circ$ toward the west direction with superconducting combined function magnet.
\item Final focus section: Match the beam optics to target both in position and in profile. Level of mm control is necessary which
corresponds to 1~mrad
$\nu$ direction difference, also not to destroy target.
\item Graphite target and horn magnet: Produce intense secondary $\pi$'s and focus them to the west direction. (3~horns system 
with 320~kA pulse operation) 
\item Muon monitor: Monitor $\mu$ direction ($=$ $\nu$ direction), pulse to pulse, with measuring center of muon profile. 
\item  On-axis neutrino monitor (INGRID): Monitor $\nu$ direction and intensity.
\end{itemize}
This facility is designed to be tolerable up to $\sim$2~MW beam power. The limitation is due to temperature rise and thermal shock
for the components such as  Al horn, graphite target, and Ti vacuum window.
Since everywhere suffers from high radiation, careful treatment of radioactive water and air ($\sim$10~GBq/3weeks) is required. 
Moreover, maintenance scenario of radio active components has to be seriously planned.

On 23rd April 2009, commissioning of the facility started with the real proton beam which was delivered by MR. 
The very first shot of the proton beam, after all the beam line magnet turned on, steered into target station and
muon monitor clearly indicated production of intense muons which certified associated neutrino production.
After 9 shots of tuning, the beam is centered on the target.
%
%
%
%
%
%
With subsequent tuning and measurement, followings are achieved.
\begin{itemize}
\item Stability of the extraction beam orbit from MR is confirmed. It is tuned within 0.3~mm in position and 0.04~mrad in direction w.r.t. 
design orbit.
\item Functionality of the superconducting combined function magnet is confirmed.
\item Beam is lead to the target center without significant beam loss. Beam trajectory is tuned within 3~mm level accuracy w.r.t. design orbit.
\item Functionality of the beam monitors (beam position, beam profile, beam intensity and beam loss) are confirmed.
\item Response function of various magnets are measured.
\item Muon signal is observed which confirms neutrino production. (Muon direction corresponds to neutrino direction and muon yield corresponds to neutrino yield.)
\item The effect of pion focusing with horn magnet is confirmed. ($\times$2 which is consistent with horn configuration at that time.)
\item The information transfer from Tokai to Kamioka  on the absolute beam time information is confirmed.
\item J-PARC neutrino facility is approved by the government on radiation safety.
\end{itemize}
%
The next running of the facility is foreseen from October 2009 with following MR intensity improvement.
Production data taking 
for neutrino experiment is foreseen to start in January 2010.

\section{T2K}
T2K~\cite{Itow:2001ee} is the first experiment with J-PARC neutrino beam. 
With the combination of unprecedented high intensity neutrino source and a well established neutrino detector, Super-Kamiokande (SK),
as a far main detector, T2K will seeks for  $\nu_\mu$ to $\nu_e$ conversion phenomenon and, as a consequence,
measures an finite value of one of the neutrino mixing angle, $\theta_{13}$, with an order of magnitude 
better sensitivity compared to the prior experiments
at an atmospheric neutrino anomaly regime.
T2K also conduct precision measurement of another neutrino mixing angle, $\theta_{23}$.
Moreover, something unexpected in neutrino physics may be revealed by T2K.

The baseline of 295~km and off-axis angle of  2.5$^\circ$ are optimized, 
1) to maximize neutrino flux at the neutrino energy of the first oscillation maximum, 
2) to avoid severe $\pi^0$ background originated from high energy neutrino interaction, and 
3) to tune neutrino energy range to be optimum for a water cherenkov detector (sub GeV energy region, low multiplicity and quasi elastic interaction dominant).

\begin{figure}[htb]
\begin{center}
\includegraphics[width=9cm,angle=-90]{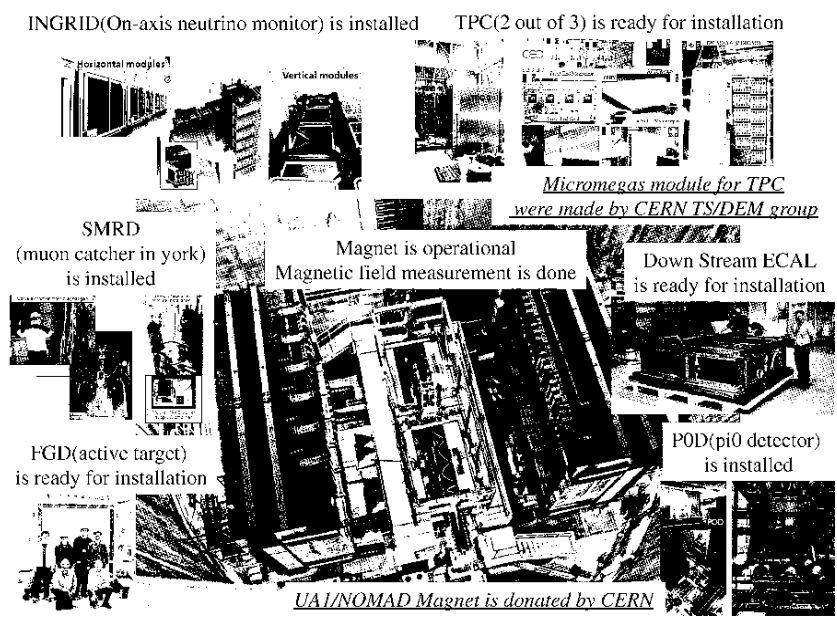}
\caption{Status of the T2K near site neutrino detectors as of October 2009}
\label{fig:ND280status}
\end{center}
\end{figure}

The properties of the produced neutrino beam are measured by a system of near detectors
at J-PARC which consists of two major parts, one is
on-axis neutrino monitor (INGRID) which 
monitors neutrino direction, intensity and its stability,  and the other is an assembly of detectors located 2.5$^\circ$ off-axis
direction as SK (ND280),
which measures not only neutrino flux as is expected at SK  but also  sub GeV neutrino interaction which gives important
information for neutrino oscillation analysis in T2K.
The most outer part of the ND280 is the UA1/NOMAD magnet,
which provides the magnetic field used to determine the momentum of charged particles originated from neutrino interaction.
Inside the magnet, Fine Grain Detector (FGD) which is an active neutrino target, Time Projection Chamber (TPC) which measures any charged particles 
emerged from neutrino interaction, $\pi^0$ detector (POD) which is optimized for measuring the rate of neutral current $\pi^0$ production, 
Electromagnetic Calorimeter (ECAL) which reconstructs any electromagnetic energy produced, and Side Muon Range Detector (SMRD)
which is instrumented in the magnet yokes to identify muons from neutrino interaction, are located.
The status of detectors as of October 2009 is  shown in Figure~\ref{fig:ND280status}.

As a first milestone, T2K 
is aiming for the first results in 2010
with 100~kw $\times$ 10$^7$~seconds integrated proton power on target
to unveil below
the CHOOZ experimental limit~\cite{Apollonio:1999ae} with $\nu_e$ appearance.

\section{European and CERN commitment on Japanese accelerator
based neutrino experiment}
\begin{figure}[htb]
\begin{center}
\includegraphics[width=6cm,angle=-90]{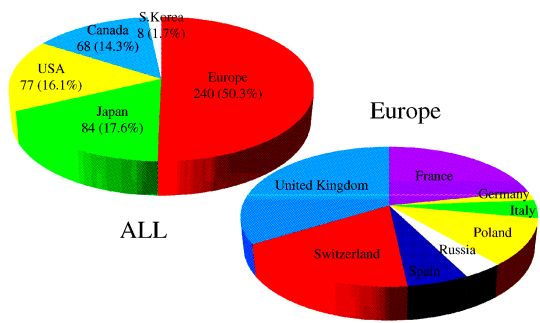}
\caption{Participants for T2K}
\label{fig:participants}
\end{center}
\end{figure}

European participation in Japanese accelerator based neutrino experiment
began at K2K (KEK to Kamioka Long Baseline Neutrino Experiment).
France, Italy, Poland, Russia, Spain and Switzerland joined this world first 
accelerator based long baseline neutrino experiment. 
When T2K project started, Germany and United Kingdom also participated. 
As of October 2009, T2K collaboration consists of 477 members from 62 institutes spread out 12 coutries.
Composition is, 240 (50.3$\%$) members from Europe, 84 (17.6$\%$) members from Japan, 77 (16.1$\%$) members from USA,
68 (14.3$\%$) members from Canada and 8 (1.7$\%$) members from South Korea, as shown in Figure~\ref{fig:participants}.
%
%
%
%
%
%

T2K is registered as Recognized Experiment at CERN (RE13) and CERN
extensively supports T2K.
Followings are the list of CERN support for T2K.
\begin{itemize}
\item CERN experiment NA61: This experiment is indispensable part of T2K to understand neutrino flux for the experiment.
\item CERN test beam for detectors.
\item Donation of UA1/NOMAD magnet.
\item Micromegas production and its test conducted by CERN TS/DEM group.
\item Various technical, administrative support on detector preparation, especially for UA1/NOMAD magnet related issues.
\item Infrastructure for detector preparation.
\item CERN-KEK cooperation on super conducting magnet for neutrino beam line.
\end{itemize}
KEK feels grateful to CERN for all the aspect of support provided by CERN.
%
%
%
%
%
%
%
%
\section{New generation accelerator based neutrino experiment in Japan}
\begin{figure}[htb]
\begin{center}
\includegraphics[width=9cm,angle=-90]{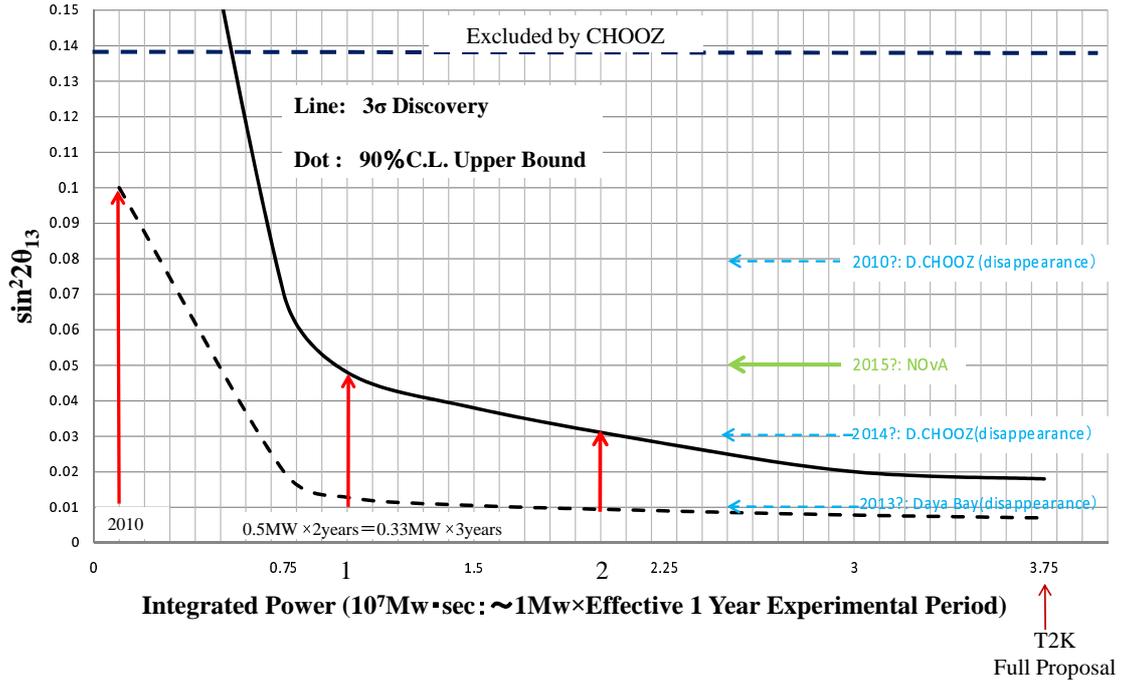}
\caption{T2K dicovery potential on $\nu_\mu\to\nu_e$ as a function of integrated proton power on target}
\label{fig:potential}
\end{center}
\end{figure}

The primary motivation of T2K
is to improve the sensitivity to the $\nu_{\mu} \rightarrow \nu_e$ conversion phenomenon 
in the atmospheric regime.
%
%
%
%
%
%
%
%
The final goal for T2K 
is to accumulate an integrated proton power on target of $0.75$~MW$\times5\times 10^7$ seconds.
As is shown in Figure~\ref{fig:potential},
%
%
within a few years of run, critical information,
which will guide the future direction of the neutrino physics, 
will be obtained based on the data corresponding to about 
1 to 2~MW$\times$10$^7$ seconds integrated proton power on target 
(roughly corresponding to a 3$\sigma$ discovery at 
sin$^2$2$\theta$$_{13}$$>$ 0.05 and 0.03, respectively). 
%
%
%
%

If a significant $\nu_\mu\to\nu_e$ coversion signal were to be observed at T2K, an immediate
step forward to a next generation experiment aimed at the discovery of CP violation
in the lepton sector would be recommended with high priority.
%
%
Compared with T2K experimental conditions, lepton sector CP violation discovery 
requires
\begin{itemize}
\item an improved J-PARC neutrino beam intensity;
\item an improved main far neutrino detector. 
\end{itemize}
Detector improvements include
\begin{itemize}
\item detector technology;
\item its volume;
\item its baseline and off-axis angle with respect to the neutrino source.
\end{itemize}
Naturally, next generation far neutrino detectors for lepton sector CP violation discovery 
will be very massive and huge. As a consequence, 
the same detector will give us the rare and important opportunity to 
discover proton decay. 
A total research subject would be, to address a long standing puzzle of our physical world,
the ``Quest for the Origin of Matter Dominated Universe" (see e.g.~\cite{cpmatter}),
%
%
%
%
%
with exploration of 
\begin{itemize}
\item the Lepton Sector CP Violation by precise testing of the neutrino oscillation processes;
\begin{itemize}
\item measure precisely the CP phase in lepton sector ($\delta$)  and the mixing angle $\theta_{13}$;
\item examine matter effect in neutrino oscillation process and possibly
conclude the mass hierarchy of neutrinos.
\end{itemize}
\item Proton Decay:
\begin{itemize}
\item Search for p $\rightarrow$ $\nu$ K$^+$ and p $\rightarrow$ e $\pi^0$
in the life time range $10^{34}$ to $10^{35}$ years,
\end{itemize}
\end{itemize}
%
with assuming non-equilibrium environment in the evolution of universe.
Even in case that sin$^2$2$\theta$$_{13}$ is below T2K sensitivity, it is still worth while trying to improve 
J-PARC neutrino beam intensity and far detector performance to open the way to explore $\nu_\mu \to \nu_e$ conversion phenomenon
with by an order of magnitude better sensitivity~\cite{okinoshima}.

This direction of research is endorsed by KEK Roadmap defined in 2008,
in which J-PARC neutrino intensity improvement and R$\&$D to realize 
huge detector for neutrino and proton decay experiments
%
are the two of the main subject.
KEK has started R$\&$D to realize huge liquid Argon time projection chamber.

%
%
%
\subsection{J-PARC neutrino beam upgrade plan}
%
%
As for the neutrino beam intensity improvement,
MR power improvement scenario toward MW-class power frontier machine,
KEK Roadmap plan, is analyzed and proposed by the J-PARC accelerator team
as shown in Table~\ref{tab:roadmap}.
\begin{table}[htb]
\begin{center}
\caption{MR power improvement scenario toward MW-class power frontier machine (KEK Roadmap)}
\label{tab:roadmap}
\begin{tabular}{lcccc}
\hline\hline
             & Start Up & Next Step & \textbf{KEK Roadmap} & Ultimate    \\
\hline
 \textbf{Power} (MW)   & 0.1 & 0.45 & \textbf{1.66}  & ? \\
 \textbf{Energy} (GeV)   & 30 & 30 & \textbf{30}  &  \\
 \textbf{Rep. Cycle} (sec.)   & 3.5 & 3-2 & \textbf{1.92}  &  \\
 \textbf{No. of Bunches}   & 6 & 8 & \textbf{8}  &  \\
 \textbf{Particles/Bunch}   & 1.2$\times$10$^{13}$ & <4.1$\times$10$^{13}$ & \textbf{8.3$\times$10$^{13}$}  &  \\
 \textbf{Particles/Ring}   & 7.2$\times$10$^{13}$ & <3.3$\times$10$^{14}$ & \textbf{6.7$\times$10$^{14}$}  &  \\
 \textbf{LINAC} (MeV)   & 181 & 181 & \textbf{400}  &  \\
 \textbf{RCS$^{a}$}   & h=2 & h=2 or 1  & \textbf{h=1}  &  \\
\hline\hline
\multicolumn{3}{l}{$^{a}$ \footnotesize Harmonic number of RCS}
\end{tabular}
\end{center}
\end{table}
%

Items to be modified from start up  toward high intensity are listed as following.
\begin{itemize}
\item Number of bunches in MR should be increased from 6 to 8.
For this purpose, fast rise time extraction kicker magnet have to be prepared.
Its installation is foreseen in 2010 summer.
\item Repetition cycle of MR has to be improved from 3.5~seconds to 1.92~seconds.
For this purpose RF and magnet power supply improvement is necessary.
\item RCS operation wih harmonic number 1 has to be conducted. This is to make the beam bunch to be longer in time domain 
to decrease space charge effect.
For this purpose RF improvement is necessary.
When RCS is operated with harmonic number 2, beam is injected to MR with 2~bunches~$\times$~4~cycles. On the other hand,
when RCS is operated with harmonic number 1, beam is injected to MR with
single bunch with doubled number of protons $\times$~8~cycles.
\item  LINAC 400~MeV operation is required  to avoid severe space charge effect at RCS injection.
Construction of necessary component is already approved and started.
\end{itemize}

\subsection{Far detector options: How to approach Lepton Sector CP Violation}
%
%
%
The effects of CP phase $\delta$ appear either 
%
\begin{itemize}
\item as a difference between $\nu$ and $\bar{\nu}$ behaviors
(this method is sensitive to the $CP$-odd term which vanishes
for $\delta=0$ or 180$^\circ$);
\item in the energy spectrum shape of the appearance oscillated $\nu_e$ charged current events 
(sensitive to all the non-vanishing $\delta$ values including 180$^\circ$).
\end{itemize}
It should be noted that
if one precisely measures the $\nu_e$  appearance energy spectrum shape (peak position and height for 1st and 2nd oscillation maximum and minimum) 
with high resolution,
CP effect could be investigated with neutrino run only.
Antineutrino beam conditions are known to be more difficult than those
for neutrinos (lower beam flux due to leading charge effect in proton collisions
on target, small antineutrinos cross-section at low energy, etc.).
%
%
%
%

\begin{figure}[htb]
\begin{center}
\includegraphics[width=6cm,angle=-90]{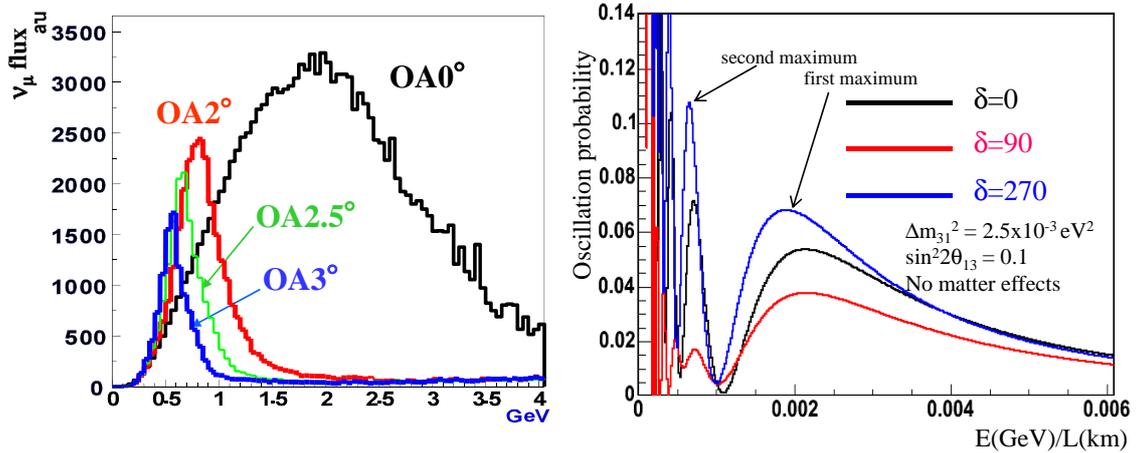}
\caption{Neutrino flux for various off-axis angle (left) and Probability for $\nu_\mu\to\nu_e$ oscillations 
as a function of the E(GeV)/L(km) for various $\delta$. (right)}
\label{fig:anglespectrum}
\end{center}
\end{figure}

Figure~\ref{fig:anglespectrum}~(left) shows neutrino flux for vaious off-axis angles.
If one selects on-axis setting, 1) wide energy coverage is foreseen which is necessary to cover the 1st and 2nd maximum simultaneously,
and  
2) measurement suffers from severe $\pi^0$ background originated from high energy neutrino which requires the detector with high performance
discrimination ability between $\pi^0$ and electron.
On the other hand, if one selects off-axis setting, 1) requirement for $\pi^0$ background discrimination is soft, and  2) measurement 
is essentially counting experiment at the 1st oscillation maximum.

Figure~\ref{fig:anglespectrum}~(right)
shows the oscillation probability as a function 
of the E(GeV)/ L(km).
If the distance between source and detector is fixed, the curves can be easily translated to
that for the expected neutrino energy spectrum of the oscillated events.
As can be seen, if the neutrino energy spectrum of the oscillated events could be reconstructed with sufficiently 
good resolution in order to distinguish first and second maximum, useful information to extract the 
CP phase would be available even only with a neutrino run.
%
%
%
If baseline is set to be long, 1) energy of 2nd oscillation maximum gets measurable.
2) statistical significance may get worse, and 3) measurement is affected by large matter effect.
On the other hand, if baseline is set to be short, 1) it is impossible to extract 2nd oscillation maximum information, 2) statistical significance may
get better, and 3) measurement is  less affected by matter effect.

To define far detector option,
discovery potential for proton decay and reality to realize huge one 
are also the essential issues to be taken into account.
%

\subsection{Possible scenarios for new generation discovery experiment with J-PARC neutrino beam}

The study of possible new generation discovery experiments with J-PARC neutrino beam was
initiated at the 4th International Workshop on Nuclear and Particle Physics at J-PARC (NP08)~\cite{NP08}.
With the same configuration as T2K (2.5$^\circ$ off-axis angle),
%
%
%
%
%
the center of the neutrino beam will go through underground beneath SK (295~km baseline), and 
will automatically reach the Okinoshima island region (658~km baseline) with an off-axis angle $0.8^\circ$ (almost on-axis) 
and eventually the sea level east of the Korean shore (1,000~km baseline) with an off-axis angle $\sim 1^\circ$. 
%
%
%
%
%
%
%

\subsubsection{Scenario 1: J-PARC to Okinoshima Long Baseline Neutrino Experiment}
\begin{figure}[htb]
\begin{center}
\includegraphics[width=9cm,angle=-90]{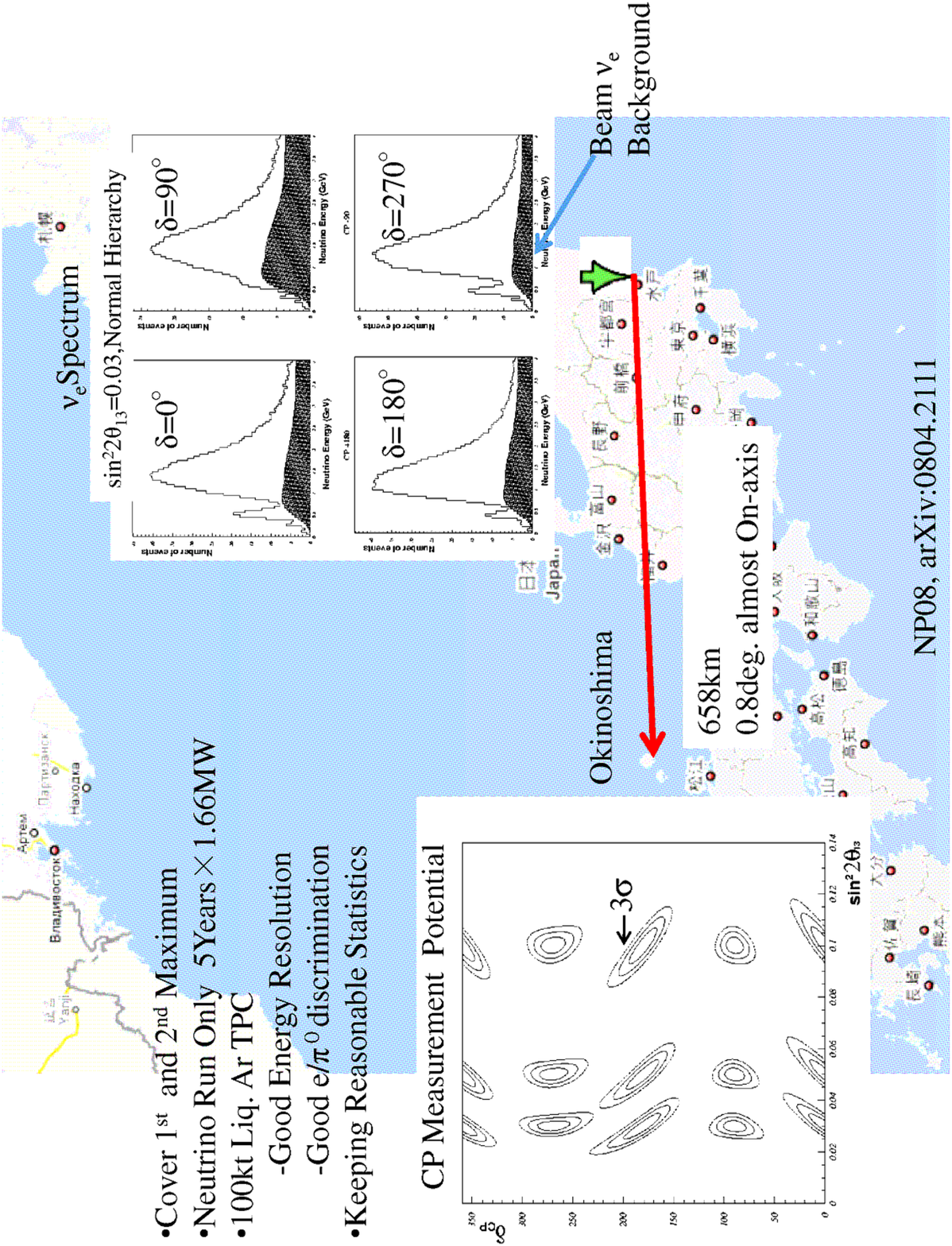}
\caption{Scenario 1: J-PARC to Okinoshima Long Baseline Neutrino Experiment}
\label{fig:scenario1}
\end{center}
\end{figure}

The first scenario is ``J-PARC to Okinoshima Long Baseline Neutrino Experiment'' as shown in Fig.~\ref{fig:scenario1}~\cite{okinoshima}.
In order to cover a wider energy range,
detector location which is near on-axis is favored.
If one assumes that the second oscillation maximum has to be located at an energy larger than  
about 400~MeV, the baseline should be longer than about 600~km.
In addition, in order to collect enough statistics, baseline should not be too much 
longer than above stated.
Taking into account all of the above mentioned considerations, 
the Okinoshima region (658~km baseline and almost on-axis (0.8$^\circ$ off-axis) configuration) turns out to be ideal.

Analysis here based on the assumption using a neutrino run only
during five years to be reasonable time duration for the single experiment ($10^7$~seconds running period/year is assumed), 
under the best J-PARC beam assumption. An anti-neutrino beam
(opposite horn polarity) might be considered in a second stage in order to cross-check
the results obtained with the neutrino run (in particular for mass hierarchy problem).
Detector is assumed to be a 100~kton liquid Argon time projection chamber.
This type of detector is supposed to provide  higher precision than other huge detectors to separate
the two peaks in energy spectrum. In addition, the $\pi^0$ background is expected to be highly suppressed
thanks to the fine granularity of the readout, hence the main irreducible background
will be the intrinsic $\nu_e$ component of the beam.
%
%
%
%
%
The right hand side plot in Figure~\ref{fig:scenario1} shows the energy spectra of electron neutrino 
at the cases 
of $\delta$ equal 0$^{\circ}$, 90$^{\circ}$, 180$^{\circ}$, 
270$^{\circ}$, respectively. Shaded region is common for all 
plots and it shows the background from beam $\nu_e$.
Here perfect resolution is assumed.
%
%
%
%
%
%
%
%
%
%
%
%
As shown, the value of $\delta$ varies the energy 
spectrum, especially the first and the second oscillation peaks (heights
and positions), therefore comparison of the peaks
determine the value $\delta$, while the value of $\sin^2 2\theta_{13}$ changes 
number of events predominantly.
Allowed regions in the perfect resolution case are shown in left hand side of 
Figure~\ref{fig:scenario1}. Twelve allowed regions are overlaid for twelve true values,
$\sin^2 2\theta_{13}$=0.1, 0.05, 0.02, and $\delta$=0$^{\circ}$,
 90$^{\circ}$, 180$^{\circ}$, 270$^{\circ}$, respectively.
The $\delta$ sensitivity is 20$\sim$30$^\circ$ depending 
on the true $\delta$ value. 


\subsubsection{Scenario 2: J-PARC to Kamioka Long Baseline Neutrino Experiment}
\begin{figure}[htb]
\begin{center}
\includegraphics[width=9cm,angle=-90]{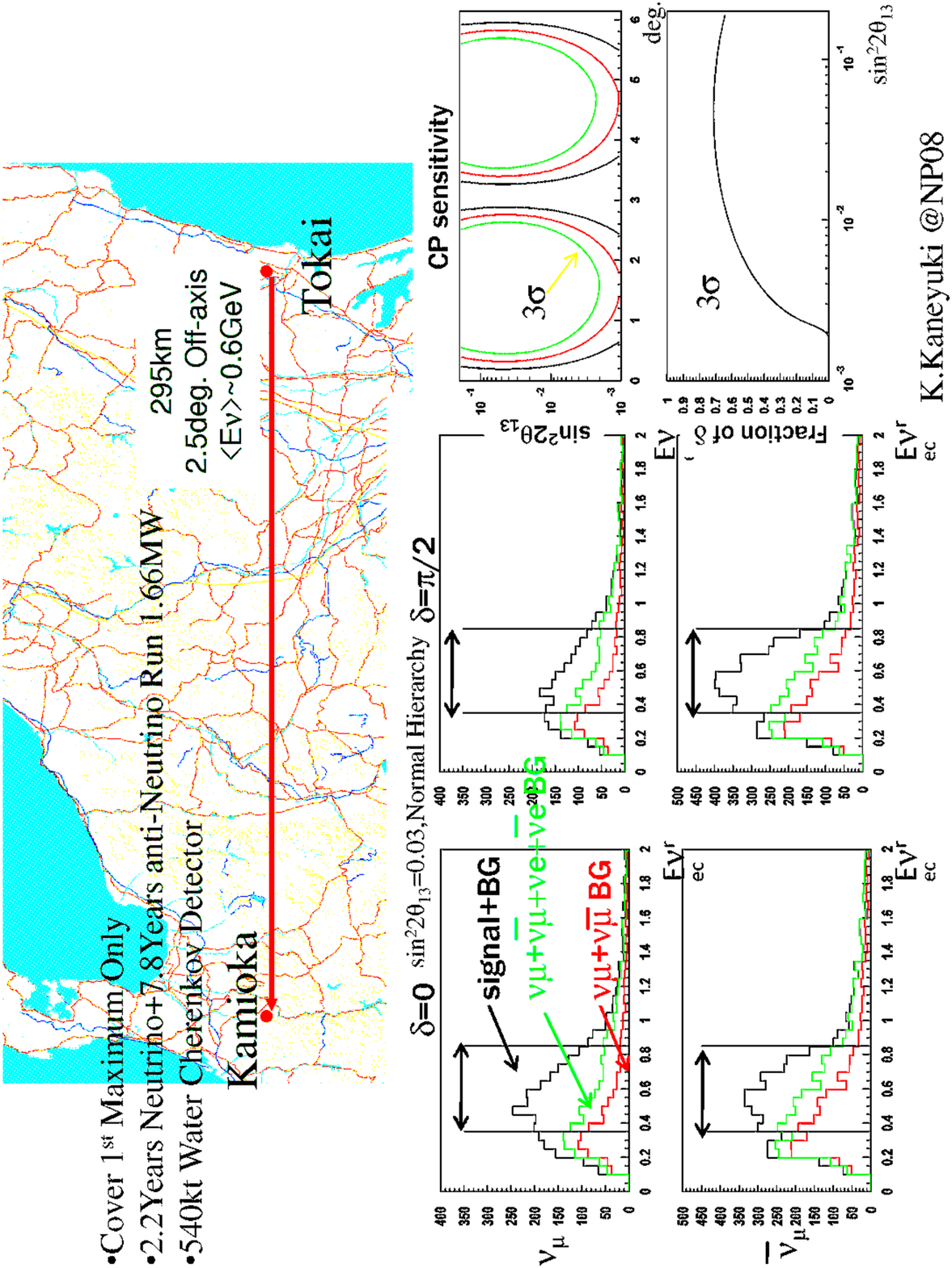}
\caption{Scenario 2: J-PARC to Kamioka Long Baseline Neutrino Experiment}
\label{fig:scenario2}
\end{center}
\end{figure}

Second scenario is ``J-PARC to Kamioka Long Baseline Neutrino Experiment'' as shown in Fig.~\ref{fig:scenario2}~\cite{NP08}.
The concept is same as T2K except for huge detector size whose fiducial volume is assumed to be 570~kt.
%
The baseline of 295~km and off-axis angle of  2.5$^\circ$ are optimum for the experimental sensitivity with a water cherenkov detector
as they are  for T2K.
%
With this configuration, on the other hand, it is only possible to cover 1st oscillation maximum.
In order to investigate difference between neutrino and anti-neutrino behavior with sufficient statistics,
2.2~years ($10^7$~seconds running period/year is assumed) neutrino run and 7.8~years anti-neutrino run is required.
Since the cancellation of systematic uncertainty between neutrino run and anti-neutrino run is not much expected,
the way to deal with delicate systematic uncertainty is a important issue to be seriously considered.

\subsubsection{Scenario3: J-PARC to Kamioka and Korea Long Baseline Neutrino Experiment}
\begin{figure}[htb]
\begin{center}
\includegraphics[width=9cm,angle=-90]{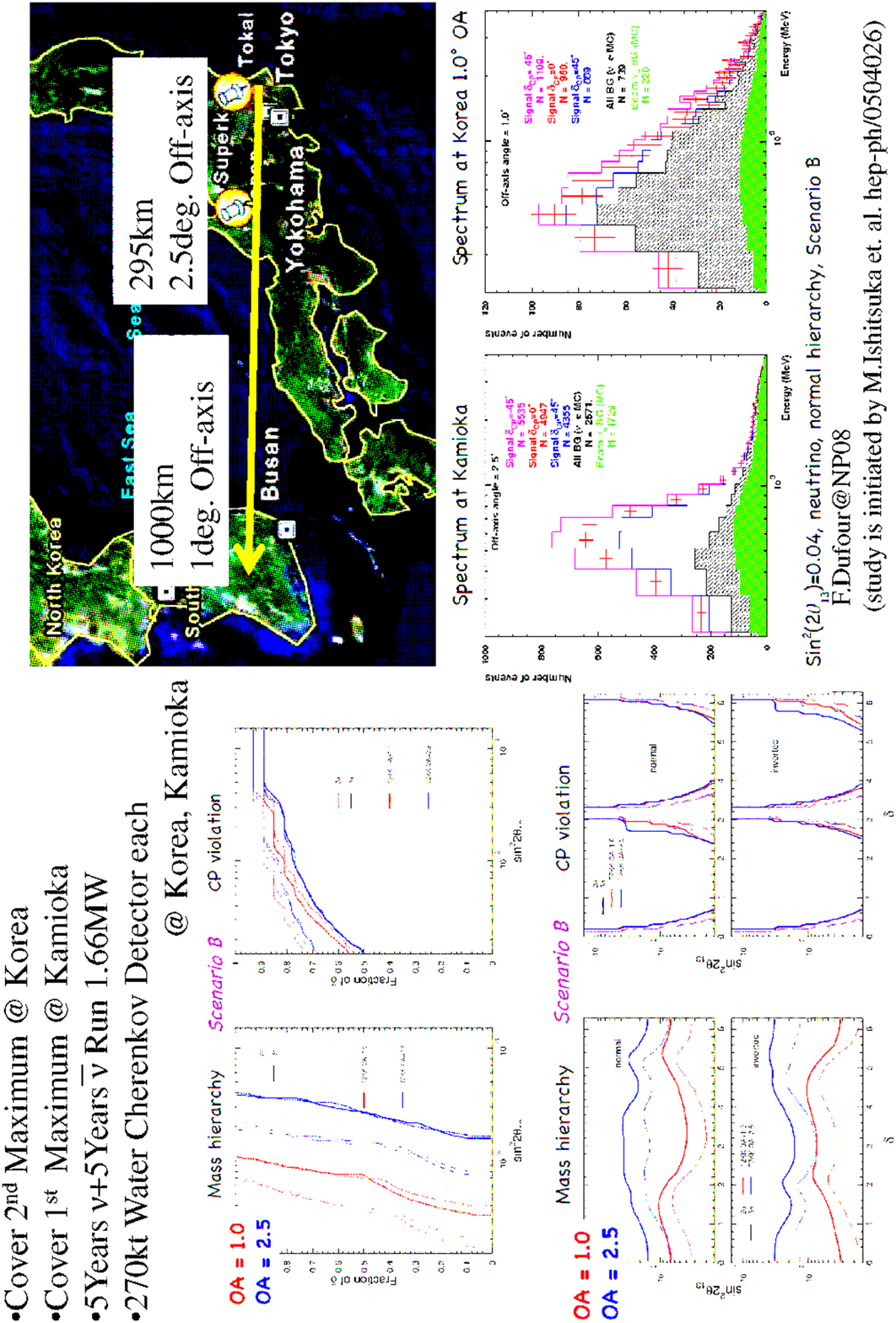}
\caption{Scenario 3: J-PARC to Kamioka and Korea Long Baseline Neutrino Experiment}
\label{fig:scenario3}
\end{center}
\end{figure}

Third scenario is ``J-PARC to Kamioka and Korea Long Baseline Neutrino Experiment'' as shown in Fig.~\ref{fig:scenario3}~\cite{NP08}.
This plan is partially same as scenario 2.
In order to obtain 1st and 2nd oscillation maxima information, in addition to neutrino and anti-neutrino difference,
two 270~kt water cherenkov detectors, one at Kamioka (295~km baseline) and the other in Korea (1,000~km baseline) are utilized.
It would allow to study E/L regions corresponding to the 1st oscillation maximum at Kamioka and 2nd oscillation maximum at Korea,
at the suitable energy regime for the measurement with a water cherenkov detector.
It requires five years each for neutrino and anti-neutrino run.

Comparison of each scenario is shown in Table~\ref{tab:scenario}. 
Study is continuing to seek for optimum choice to maximize potential of the research.
%
%
\begin{table}[htb]
\begin{center}
\caption{Comparison of possible scenarios for new generation discovery experiment with J-PARC neutrino beam}
\label{tab:scenario}
\begin{tabular}{lccc}
\hline\hline
\textbf{}             & \textbf{Scenario 1} & \textbf{Scenario 2} & \textbf{Scenario 3} \\
\textbf{}             & \textbf{Okinoshima} & \textbf{Kamioka} & \textbf{Kamioka and Korea} \\
\hline
 \textbf{Baseline} (km)  & 658 & 295 & 295 and 1000  \\
 \textbf{Off-Axis Angle} ($^\circ$)  & 0.8(almost on-axis) & 2.5 & 2.5 and 1 \\
 \textbf{Method}   & $\nu_e$ Spectrum Shape & Ratio between $\nu_e$ and $\bar{\nu}_e$ &  Ratio between 1st and 2nd Max.   \\
    &  &  &  Ratio between $\nu_e$ and $\bar{\nu}_e$  \\
 \textbf{Beam}   & 5 years $\nu_\mu$ & 2.2 years $\nu_\mu$  &  5 years $\nu_\mu$ \\
    & then Decide Next &  and 7.8 years $\bar{\nu}_\mu$ &  and 5 years $\bar{\nu}_\mu$ \\
 \textbf{Detector Technology}   & Liq. Ar TPC & Water Cherenkov & Water Cherenkov \\
 \textbf{Detector Mass} (kt)   & 100 & 2$\times$270 & 270+270 \\ 
\hline\hline
\end{tabular}
\end{center}
\end{table}
%
%

\section{Accelerator based neutrino project in Japan}

Table~\ref{tab:project} summarizes accelerator based neutrino project in Japan.
When K2K and T2K projects started, the existence of high performance far main detector, SK, 
made it possible to concentrate on neutrino beam source related preparation.
As for the 3rd generation experiment, the existence of J-PARC neutrino beam make it possible to concentrate on 
far detector issues after T2K starts up.
%
%
\begin{table}[htb]
\begin{center}

\caption{Accelerator based neutrino project in Japan}
\label{tab:project}
\begin{tabular}{llll}
\hline\hline
\textbf{}             & K2K & \ T2K & \textbf{3rd Generation Experiment} \\
\hline
 \textbf{High Power}   & KEK PS & J-PARC MR & \textbf{J-PARC MR}  \\
 \textbf{Proton}   & 12GeV 0.005MW & 30GeV 0.75MW & \textbf{30GeV 1.66MW}  \\
 \textbf{Synchrotron}   & Existing & Brand New & \textbf{Technically Feasible Upgrade}  \\
\   &  &  &   \\
\textbf{Neutrino Beamline}   & K2K & J-PARC & \textbf{J-PARC}  \\
\textbf{}   & Neutrino Beamline & Neutrino Beamline & \textbf{Neutrino Beamline}  \\
\textbf{}   & Brand New & Brand New & \textbf{Existing}  \\
\   &  &  &   \\
\textbf{Far Detector}   & Super Kamiokande & Super Kamiokand & \textbf{Brand New}  \\
\textbf{}   & Existing at  & Existing at  & \textbf{- Detector Technology ?}  \\
\textbf{}   & KAMIOKA  & KAMIOKA  & \textbf{- Place (Angle and Baseline) ?}  \\
\   &  &  &   \\
\textbf{1st Priority Physics Case}   & Neutrino Oscillation  & Neutrino Oscillation & \textbf{Lepton Sector CP Violation}  \\
\textbf{}   & $\nu_\mu$ Disappearance  & $\nu_\mu \to \nu_e$ & \textbf{Proton Decay}  \\

\hline\hline
\end{tabular}
\end{center}
\end{table}

To complete present project T2K successfully and realize new generation discovery experiment,
following issues are important.
\begin{itemize}
\item Deliver high quality experimental output from T2K as soon as possible. 
\item Realize quick improvement of accelerator power toward MW-class power frontier machine.
\item Validate beam line components tolerance (especially pion production target related issues) toward MW proton beam. 
\item  Conduct intensive R$\&$D on realization of huge liquid Argon time projection chamber and water cherenkov detector. 
\end{itemize}
Healthy scientific competition and cooperation in the world is key to promote high energy physics.
It is welcomed to cooperate in any aspects.
%
%

In Japan we will proceed as following.
\begin{itemize}
\item Short term
\begin{itemize}
\item Beam commissioning of J-PARC MR has started May-2008.
\item Commissioning of J-PARC neutrino beam facility has started in April-2009.
\item T2K is aiming for the first results in 2010 with 100~kw~$\times$~10$^7$~seconds integrated proton power 
on target to unveil below CHOOZ experimental limit with $\nu_e$ appearance. 
\end{itemize}
\item Middle term
\begin{itemize}
\item T2K data with 1-2~MW~$\times$~10$^7$~seconds integrated proton power on target will provide critical information on
 $\theta_{13}$, which guides the future direction of the neutrino physics. (In any case, complete T2K proposal 
of 3.75~MW~$\times$~10$^7$~seconds.)
\item Achieve MR power improvement scenario toward MW-class power frontier machine (KEK Roadmap).
\item  Submit proposal "J-PARC to Somewhere Long Baseline Neutrino Experiment and 
Nucleon Decay Experiment with Huge Detector" and construct huge detector. 
\end{itemize}
\item Long term
\begin{itemize}
\item Discover CP violation in lepton sector and  proton decay, and solve ``Quest for the Origin of Matter Dominated Universe"
\end{itemize}
\end{itemize}
\end{document}